\documentclass{revtex4}
\usepackage{amssymb}
\usepackage{amsmath}

\begin{document}

\title{The optical model of $N\overline{N}$ interaction without cut-off radius.}
\author{O.D. Dalkarov, A. Yu. Voronin\\P. N. Lebedev Physical Institute\\53 Leninsky pr.,117924 Moscow, Russia}

\begin{abstract}
We suggest a regular potential model of $N\overline{N}$ interaction without
any cut-off. The effect of singular terms of OBE potential, modified by
annihilation is shown to be repulsive. The experimental data for S- and P-wave
scattering lengths are well reproduced.

\end{abstract}
\maketitle\maketitle
\bigskip

\bigskip

\section{Introduction}

During the last decades numerous nonrelativistic models of
$N\overline{N}$ low energy interaction
\cite{Phil,DM,CPS,KW,DR,Pign} have been suggested as well as great
experimental efforts have been made on the Low Energy Antiproton
Ring (LEAR) at CERN. An intriguing problem of possible existing of
quasi-nuclear $N\overline{N}$ states \cite{Shapiro} strongly
stimulated the mentioned researches. The physical arguments in
favor of quasi-nuclear states are the following. The interaction
between $N$ and $\overline{N}$ should be much more attractive than
the $NN$ interaction, as it follows from the procedure of
$G$-conjugation \cite{Phil}. Such a strong attraction should
produce a spectrum of $N\overline{N}$ quasi-bound states (so
called \textit{baryonium}). In the same time the range of
annihilation, estimated from the position of the nearest to the
threshold singularity in Feynman annihilation diagrams, is much
smaller than the range of meson exchange forces. This means that
baryonium states could be rather narrow to be observed
experimentally. It was indeed discovered in LEAR experiments that
certain partial waves contain narrow nearthreshold resonances (so
called $P$-wave enhancement). Unfortunately, this transparent
physical picture has a significant drawback. The $G$-conjugation
of $NN$ OBE potential yields in attractive singular terms in
$N\overline{N}$ potential of the type $1/r^{3}$. (In case of $NN$
these terms are repulsive and play a role of so called short range
core). It is well known, that attractive singular potentials
produce a collapse \cite{LL}, i.e. the spectrum of the system is
not bounded from below, while the scattering problem has no
definite solution. The usual way of dealing with such pathological
potentials is to impose, that singular behavior is an artifact of
certain approximations (for instance nonrelativistic
approximation). In the absence of the self-consistent theory it is
common practice to introduce the phenomenological cut-off radius
to regularize the singular behavior of the model at short
distance. However the results change dramatically with small
variations of the cut-off radius\cite{Carb}, which seriously
diminish predictive power of the model. In fact it is not clear if
the nearthreshold states are determined by the ''physical part''
of the OBEP, or they are artifacts, produced by the
''non-physical'', singular part of the interaction.

The aim of the present study is to analyze the role of singular
interaction and suggest a model of $N\overline{N}$ interaction
which is free from the non-physical cut-off parameters. The main
idea of our approach is that strong enough short range
annihilation regularizes the singular attractive potentials, as
far as the particle annihilates rather than falls to the center.
Mathematically this means that singular potential can be
regularized by \textit{complex }addition to the coupling constant.
The main properties of the potentials of the type
$-(\alpha+i\omega)/r^{s}$, $s\geq2$ were studied in \cite{AV}. It
was shown that such potentials with $\omega\neq0$ become regular
and the scattering on such a potential is equivalent to the full
absorption of the particles in the scattering center. These
results enable to suggest a regular potential model of
$N\overline{N}$ interaction, based on OBE potential, but without
any cut-off radius. We will show that the overall effect of the
''singular'' part of potential, modified by annihilation is
repulsive and thus cannot be responsible for new ''false'' states.
The neathreshold resonances, which are well reproduced by our
model, are determined mainly by long range part of OBEP. We
calculate the scattering lengths in S- and P- partial waves for
different values of spin, isospin and total momentum and
demonstrate that our model reproduces rather well the experimental
data.

\section{Singular potentials with complex strength.}

  In this section we present the main results concerning the
properties of singular potentials with a complex strength. In the
following we put $2M=1$ and consider the interaction strength to
be complex $\alpha _{s}=\mathop{\rm Re}\alpha _{s}\pm i\omega $.
Near the origin we can neglect  all the nonsingular terms of the
Shr\"{o}dinger equation, increasing slower than $1/r^{2}$. One get
the following expression for the wave-function near the origin
\cite{MM}:

\begin{eqnarray}
\Phi (r) &=&\sqrt{r}\left(H_{\mu }^{(1)}(z)+\exp
(2i\delta_0)H_{\mu}^{(2)}(z)\right)\label{Phi}\\
z &=&\frac{2\sqrt{\alpha _{s}}}{s-2}r^{-(s-2)/2}\\
\mu &=&\frac{2l+1}{s-2}
\end{eqnarray}

Here $H_{\mu }^{(1)}(z)$ and $H_{\mu }^{(2)}(z)$ are the Hankel
functions of order $\mu $ \cite{Watson}, $\delta_0$ is a
contribution of the short range part of the singular potential
into the scattering phase. It is worth to mention that the
variable $z$ is a semiclassical phase.

Let us replace the singular potential at distance less than
$r_{0}$ by the constant  $-\alpha/r_{0}^s$, having in mind to tend
$r_0\rightarrow 0$. Matching the logarithmic derivatives for the
"square-well" solution and the solution (\ref{Phi}) at small $r_0$
, one can get for $\delta_0$:
\begin{eqnarray}
\delta_0&=&p(r_0)r_0 \label{delta0} \\
p(r_0)&=&\frac{\sqrt{\mathop{\rm Re}\alpha _{s}\pm i\omega
}}{r_0^{s/2}}\label{peff}
\end{eqnarray}

Now it is important that the interaction strength $\alpha_s$ is
complex. In the limit $r_0\rightarrow 0$ we obtain:

\begin{equation}
\lim_{r_0\rightarrow 0}\mathop{\rm Im}\delta _{0}=\mathop{\rm
Im}\frac{\sqrt{\mathop{\rm Re}\alpha _{s}\pm i\omega
}}{r_0^{(s-2)/2}}\rightarrow \pm \infty \label{Limdelt}
\end{equation}
which means, that $\exp(2i\delta_0)$ is either $0$ or $\infty$ and
the linear combination of the Shr\"{o}dinger equation solutions
(\ref{Phi}) is uniquely defined in the limit of zero cut-off
radius $r_0$:
\begin{equation}
\lim_{r_0\rightarrow 0}\Phi(r)=\left\{
\begin{array}{cll}
\sqrt{r}H_{\mu }^{(1)}(z) & \mbox{if} & \omega > 0 \\
\sqrt{r}H_{\mu }^{(2)}(z) & \mbox{if} & \omega < 0%
\end{array}
\right.  \label{bound}
\end{equation}
One can see, that $\omega > 0$ selects an incoming wave , which
corresponds to the full absorption of the particle in the
scattering center, while $\omega < 0$ selects an outgoing wave,
which corresponds to the creation of the particle in the
scattering center.

As one can see from (\ref{delta0}) and (\ref{peff}) due to the
singular character of our potential ($s>2$), the above conclusions
are valid for any infinitesimal value of $\omega$. It means, that
the sign of an infinitesimal imaginary addition to the interaction
constant selects the full absorption or the full creation boundary
condition (\ref{bound}). This boundary condition can be formulated
in terms of the logarithmic derivative in the origin:
\begin{equation}
\lim_{r\rightarrow 0}\frac{\Phi'(r)}{\Phi(r)}=- i \mathop{\rm
sign}(\omega) p(r)
\end{equation}
where $p(r)$ is a classical local momentum (\ref{peff}). (Compare
with plane incoming (outgoing) wave boundary condition $\exp(\mp
ipr)'/\exp(\mp ipr)=\mp i p$).

As soon as solution of the Shr\"{o}dinger equation is uniquely
defined, we can  calculate the scattering observables. In
particular we can now obtain the S-wave scattering length for the
potential $-(\alpha _{s}\pm i0)/r^{s}$ (for $s>3$) :

\begin{equation}
a=\exp (\mp i\pi /(s-2))\left( \frac{\sqrt{\alpha
_{s}}}{s-2}\right) ^{2/(s-2)}\frac{\Gamma ((s-3)/(s-2))}{\Gamma
((s-1)/(s-2))}\label{sclength}
\end{equation}

The fact, that in spite $\mathop{\rm Im}\alpha _{s}\rightarrow \pm
0$ the scattering length has nonzero imaginary part is the
manifestation of the singular properties of the potential which
violates the self-adjointness of the Hamiltonian.

 Let us compare the scattering length (\ref%
{sclength}) with that of the repulsive singular potential $\alpha _{s}/r^{s}$%
. One can get:
\begin{equation}
a^{rep}=\left( \frac{\sqrt{\alpha _{s}}}{s-2}\right)
^{2/(s-2)}\frac{\Gamma ((s-3)/(s-2))}{\Gamma ((s-1)/(s-2))}
\label{screpulsive}
\end{equation}

It is easy to see, that (\ref{sclength}) can be obtained from (\ref%
{screpulsive}) simply by choosing the certain branch
(corresponding to an absorption or a creation) of the function
$\left( \sqrt{\alpha _{s}}\right) ^{2/(s-2)}$ when passing through
the branching point $\alpha _{s}=0$. The scattering length in a
regularized singular potential becomes an analytical function of
$\alpha _{s}$ in the whole complex  plane of $\alpha _{s}$ with a
cut along positive real axis. One can see, that the presence of an
inelastic component in the singular potential acts in the same
way, as a repulsion. It suppresses one of two solutions of the
Schr\"{o}dinger equation and thus eliminates the collapse.

It is easy to see, that the boundary condition (\ref{bound}) of
the full absorption (creation) is incompatible with the existence
of any bound state. Indeed, one needs both incoming and reflected
wave to form a standing wave, corresponding to a bound state. This
means, that the regularized singular potential supports \textbf{no
bound states}. This is also clear from the mentioned above fact,
that the scattering length for a regularized attractive singular
potential is an analytical continuation of the scattering length
of a repulsive potential.

\subsection{Potential $-\protect\alpha _{2}/r^{2}$}

Let us now turn to the very important case $-\alpha /r^{2}$. The
wave-function now is:
\begin{eqnarray}
\Phi  &=&\sqrt{r}\left[ J_{\nu _{+}}(kr)+\exp
(2i\delta_0)J_{\nu _{-}}(kr)\right]  \\
\nu _{\pm } &=&\pm \sqrt{1/4-\alpha _{2}}
\end{eqnarray}
where $k=\sqrt{E}$, and $J_{\nu _{\pm }}$ are the Bessel functions \cite%
{Watson}. In the following we will be interested in the values of
$\mathop{\rm Re}\alpha _{2}$ greater than critical $\mathop{\rm
Re}\alpha _{2}>1/4$. We use the same cut-off procedure at small
$r_0$.
Matching the logarithmic derivatives at cut-off point $r_0$ we get for $%
\exp (2i\delta_0)$:
\[
\lim_{r_0\rightarrow 0}\exp (2i\delta_0)=r_{0}^{\nu _{+}-\nu
_{-}}const\sim r_{0}^{2\nu }=r_{0}^{\omega /\sqrt{\mathop{\rm
Re}\alpha _{2}-1/4}}r_{0}^{-2i\sqrt{\mathop{\rm Re}\alpha
_{2}-1/4}}
\]

One can see, that due to the presence of an imaginary addition $
\omega$ we get $\mathop{\rm Im}\delta _{0}\rightarrow \pm \infty$
when $r_{0}\rightarrow 0$.

Again we come to the boundary condition:
\begin{equation}
\lim_{r_0\rightarrow 0}\Phi(r)=\left\{
\begin{array}{cll}
\sqrt{r}J_{\nu _{+}}(kr) & \mbox{if} & \omega > 0 \\
\sqrt{r}J_{\nu _{-}}(kr) & \mbox{if} & \omega < 0%
\end{array}
\right.  \label{bound2}
\end{equation}
where $\nu _{_{\pm}}=\pm\sqrt{1/4-\alpha _{2}}$

For the large argument this function behaves like:
\[
\Phi \sim \cos (z-\nu _{_{\pm}}\pi /2-\pi /4)
\]
The corresponding scattering phase is:
\begin{equation}
\delta =\frac{\pm i\pi }{2}\sqrt{\alpha _{2}-1/4}+\pi /4
\label{scphase}
\end{equation}

As one can see, the S-matrix $S=\exp(2i\delta)$ is energy
independent. This means that the regularized inverse square
potential supports \textbf{no bound states}. The regularized
wave-function and the phase-shift are analytical functions of
$\alpha _{2}$ in the whole complex plane with a cut along the axis
$\mathop{\rm Re}\alpha _{2}>1/4.$

\subsection{Critical strength of inelastic interaction}
Now we would like to determine "how strong" should be annihilation
to regularize the attractive singular potential of order $s$. In
other words we would like to find the minimum order of singularity
of an infinitesimal imaginary potential required for the
regularization of given singular potential. The potential of
interest is a sum $-\alpha _{s}/r^{s}\mp i\omega/r^{t}$. Here we
keep $\alpha _{s}$ real. From expressions (\ref{delta0},
\ref{peff}) one immediately comes to the conclusion that the
regularization is possible only if $t>s/2+1$.

 The above statement makes clear the physical sense of suggested regularization.
 The scattering is insensitive to any details of the short range modification
 of a
singular interaction if the inelastic component of such an
interaction behaves more singular than $-1/r^{(s/2+1)}$.

\subsection{Singular potential and WKB approximation.}
The WKB approximation  holds if $|\frac{\partial (1/p)}{\partial
r}|\ll 1$. In case of the zero-energy scattering on a singular
potential with $s>2$ this condition is valid for $r\ll
r_{sc}\equiv (2\sqrt{\alpha _{s}}/s)^{2/(s-2)}$ , i.e. near the
origin. (For $s=2$ the semiclassical approximation is valid only
for $\alpha \gg 1$).
The WKB approximation, consistent with the boundary condition (\ref{bound}) for $%
s>2 $ is :
\begin{equation}
\Phi =\frac{1}{\sqrt{p(r)}}\exp (\pm i\int\limits_{r}^{a}p(r)dr)
\label{semicl}
\end{equation}
with $p(r)$ from (\ref{peff}). It follows from the above
expression, that in case the WKB approximation is valid everywhere
the solution of the Schr\"{o}dinger equation  includes incoming
wave only (for distinctness we speak here of absorptive
potential). The corresponding S-matrix is equal to zero $S=0$
within such an approximation and insensitive to any modifications
of the inner part of potential $p^2(r)$. The outgoing wave can
appear in the solution only in the regions where (\ref{semicl})
does not hold. For example, in the zero energy limit
$k^2\rightarrow 0$ the S-matrix is nonzero $S=1-2ika$. One can
show that the outgoing wave is reflected from those parts of the
potential which change sufficiently fast in comparison with the
effective wavelength (so called quantum
reflection)$|\frac{\partial (1/p)}{\partial r}|\gtrsim 1$

For the zero energy scattering and $l=0$ this holds for $r\geq
r_{sc}$.

The reflection coefficient $P\equiv|S|^2=\exp(-4\mathop{\rm
Im}\delta)$ which shows the reflected part of the flux  has the
following form in the low energy limit:
\[
P=1-4 k\mathop{\rm Im}a
\]
For the energies $E\gg E_{sc}\equiv(s/2)^{2s/(s-2)}\alpha
_{s}^{-2/(s-2)}$ the WKB holds everywhere and S-wave reflection
becomes exponentially small.

 An important conclusion is that any
information, which comes from the scattering on an absorptive
singular potential is due to a quantum reflection from the tail of
such a potential.

\section{Optical model of $N\overline{N}$ interaction.}

From the above results it is clear that the model potential, which
behaves at short distance like $-(\alpha+i\omega)/r^{3}$ is
regular, i.e. it enables definite unique solution of the
scattering problem. Such a potential is absorptive and describes
not only elastic, but inelastic scattering as well. We suggest the
following model potential:
\begin{equation}
W=V_{KW}-i\frac{A}{r^{3}}\exp(-r/\tau)\label{MP}%
\end{equation}
here $V_{KW}$ is the real part of Kohno-Weise version of OBE
potential\cite{KW}, but without any cut-off. The parameters of
imaginary part of the potential were choosed as follows: A=4.7
$GeV$ $fm^{2}$, $\tau=0.4$ $fm$. We have calculated the values of
S- and P-scattering lengths in our model potential. The obtained
results, (indicated as Reg) together with the results of two
Dover-Richard models, (DR1 and DR2), and
Kohno-Weise model (KW) are presented in the table. \bigskip%

\begin{tabular}{|c|c|c|c|c|}
\hline
 State & DR1 & DR2 & KW & Reg\\ \hline
$^{11}S_{0}$ & 0.02-i1.12 & 0.1-i1.06 & -0.03-i1.35 &
-0.08-i1.16\\ \hline
$^{31}S_{0}$ & 1.17-i0.51 & 1.2-i0.57 &
1.07-i0.62 & 1.05-i0.55\\ \hline
$^{13}S_{1}$ & 1.16-i0.46 &
1.16-i0.47 & 1.24-i0.63 & 1.19-i0.64\\ \hline
$^{33}S_{1}$ &
0.86-i0.63 & 0.87-i0.67 & 0.71-i0.76 & 0.7-i0.65\\ \hline
$^{11}P_{1}$ & -3.33-i0.56 & -3.28-i0.78 & -3.36-i0.62 &
-3.19-i0.59\\ \hline
$^{31}P_{1}$ & 0.92-i0.5 & 1.02-i0.43 &
0.71-i0.47 & 0.81-i0.46\\ \hline
$^{13}P_{0}$ & -9.58-i5.2 &
-8.53-i3.51 & -8.83-i4.45 & -7.67-i4.74\\ \hline
$^{33}P_{0}$ &
2.69-i0.13 & 2.67-i0.15 & 2.43-i0.11 & 2.46-i0.15\\ \hline
$^{13}P_{1}$ & 5.16-i0.08 & 5.14-i0.09 & 4.73-i0.08 & 4.75-i0.15\\
\hline $^{33}P_{1}$ & -2.08-i0.86 & -2.02-i0.7 & -2.17-i0.95 &
-2.09-i0.79\\ \hline $^{13}P_{2}$ & 0.04-i0.57 & 0.22-i0.56 &
-0.03-i0.88 & -0.12-i0.82\\ \hline $^{33}P_{2}$ & -0.1-i0.46 &
0.05-i0.55 & -0.25-i0.39 & -0.14-i0.39\\ \hline
\end{tabular}

\bigskip
One can see rather good agreement between the results obtained
within the  suggested   optical model without cut-off and cited
above versions of Kohno-Weise and Dover-Richard models.
\section{Conclusion}
We have found that the scattering observables are insensitive to
any details of the short range modification of a singular
interaction, if such an interaction includes strong inelastic
component. In this case the scattering amplitude can be calculated
by
solving the Schr\"{o}dinger equation with the regularized singular potential $%
-(\alpha _{s}\pm i0)/r^{s}$. It was shown that  the low energy
scattering amplitude on such a potential is determined by the
quantum scattering from the region, where WKB approximation fails
( the potential tail). The mentioned formalism was used to build
an optical model of $N\bar{N}$ low energy interaction free from
uncertainty, related to the cut-off parameter. The good agreement
of the results, obtained within our regularization method and
within different $N\bar{N}$ interaction models proves that the
long range part (pionic tail) of OBEP plays essential role for the
nearthreshold scattering.

\section{Acknowledgement}

The research was performed under support of Russian Foundation for
Basic Research grant 02-02-16809.

\end{document}